\documentclass[sigconf]{acmart}

\usepackage{balance}

\usepackage{booktabs}
\usepackage{multicol}
\usepackage{multirow}
\usepackage{amssymb}
\usepackage{amsmath}

\def\BibTeX{{\rm B\kern-.05em{\sc i\kern-.025em b}\kern-.08emT\kern-.1667em\lower.7ex\hbox{E}\kern-.125emX}}
    
\copyrightyear{2019} 
\acmYear{2019} 
\setcopyright{acmcopyright}
\acmConference[KDD '19]{The 25th ACM SIGKDD Conference on Knowledge Discovery and Data Mining}{August 4--8, 2019}{Anchorage, AK, USA}
\acmBooktitle{The 25th ACM SIGKDD Conference on Knowledge Discovery and Data Mining (KDD '19), August 4--8, 2019, Anchorage, AK, USA}
\acmPrice{15.00}
\acmDOI{10.1145/3292500.3330931}
\acmISBN{978-1-4503-6201-6/19/08}

\begin{document}

\title{MCNE: An End-to-End Framework for Learning Multiple Conditional Network Representations of Social Network}


\author{Hao Wang$^{1}$, Tong Xu$^{1, *}$, Qi Liu$^{1}$, Defu Lian$^{1}$, Enhong Chen$^{1, *}$, Dongfang Du$^{2}$, Han Wu$^{1}$, Wen Su$^{2}$}
\affiliation{%
	\institution{$^1$Anhui Province Key Lab of Big Data Analysis and Application, School of Computer Science and Technology,\\
		University of Science and Technology of China\\
		$^{2}$Tencent Inc, China}}
\affiliation{\{wanghao3, wuhanhan\}@mail.ustc.edu.cn, \{tongxu, qiliuql, liandefu, cheneh\}@ustc.edu.cn}
\affiliation{\{blithedu, oliviayuisu\}@tencent.com}


%
\renewcommand{\shortauthors}{Hao Wang, et al.}
\renewcommand{\shorttitle}{MCNE: Learning Multiple Conditional Network Representations}
%
\begin{abstract}
Recently, the Network Representation Learning (NRL) techniques, which represent graph structure via low-dimension vectors to support social-oriented application, have attracted wide attention. Though large efforts have been made, they may fail to describe the multiple aspects of similarity between social users, as only a single vector for one unique aspect has been represented for each node. To that end, in this paper, we propose a novel end-to-end framework named MCNE to learn multiple conditional network representations, so that various preferences for multiple behaviors could be fully captured. Specifically, we first design a binary mask layer to divide the single vector as conditional embeddings for multiple behaviors. Then, we introduce the attention network to model interaction relationship among multiple preferences, and further utilize the adapted message sending and receiving operation of graph neural network, so that multi-aspect preference information from high-order neighbors will be captured. Finally, we utilize Bayesian Personalized Ranking loss function to learn the preference similarity on each behavior, and jointly learn multiple conditional node embeddings via multi-task learning framework. Extensive experiments on public datasets validate that our MCNE framework could significantly outperform several state-of-the-art baselines, and further support the visualization and transfer learning tasks with excellent interpretability and robustness.

\end{abstract}

%
%

%
\keywords{Network Embedding, Social Netowrk, Conditional Representation\let\thefootnote\relax\footnotetext{$^*$Corresponding Author.}}

%
\maketitle

\section{Introduction}
With the development of embedding techniques, a series of Network Representation Learning (NRL) algorithms have been proposed, which aim to embed the network structure into a low-dimensional space and obtain the vector representation for each node. Then these learned node embeddings can be utilized as the features of nodes, and be directly applied to various network-oriented applications, such as node classification \cite{sen2008collective, wang2018united}, link prediction\cite{du2017solving}, node recommendation \cite{ying2018graph} and social influence analysis \cite{liu2017influence,zhang2015modeling,xiao2018price}.
\begin{figure}[t]
	\centering	
	\includegraphics[width=0.5\textwidth]{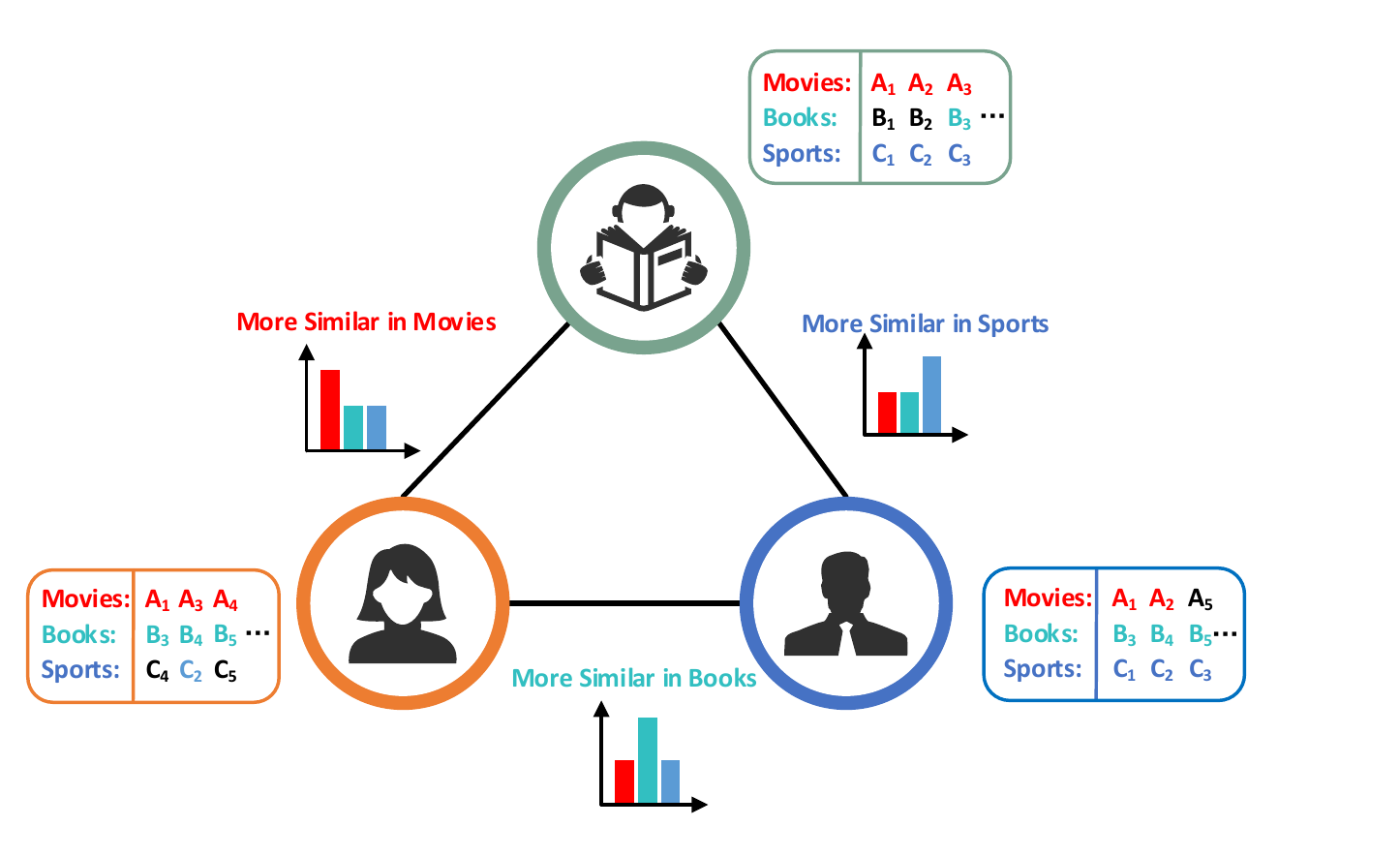}\vspace{-0.6cm}	
	\caption{Multiple Similarity Relationships of Social Users}\vspace{-0.0cm}
	\label{introduction:fig1} \vspace{-0.4cm}
\end{figure}

In the field of network representation learning, DeepWalk \cite{perozzi2014deepwalk} is the first to adopt skip-gram model proposed in Word2vec \cite{mikolov2013efficient} to learn the vector representation of nodes. Then Node2vec \cite{grover2016node2vec} further leverages two parameters to control the breadth-first sampling and depth-first sampling, which corresponds to the homogeneity and structural equivalence respectively. At the same time, LINE \cite{tang2015line} analyzes the local and global network structure, and employs the first-order and second-order proximity to derive the objective functions. SDNE \cite{wang2016structural} further utilizes the semi-supervised auto-encoder model to map the network to a highly non-linear latent space, in order to preserve the network structure and be robust to sparse networks. Furthermore, some research work combine the network representation learning methods with the specific task to further improve the performance. TriNRL \cite{pan2016tri} combines the network structure with node attributes and labels, and jointly models the inter-node relationship, node-word correlation and label-word correspondence. And GCN \cite{kipf2016semi} utilizes an approximate graph convolution operation to generate the node representation, and learns the node embedding in a semi-supervised learning graph framework. PinSage \cite{ying2018graph} is the state-of-the-art method that designs the effective graph convolutional architectures to generate the embedding of graph-structured items, and uses the max-margin-based loss function to learn the similarity between items. And it's worth noting that these network representation learning methods only learn a single vector representation for each node. This means these node embedding can only represent one unique aspect of similarity between users. However, in a real social network, there are often multiple aspects of similarity between users, which is illustrated in Figure~\ref{introduction:fig1}. From the figure, we can observe that user's preferences of different behaviors are not identical. For example, the `woman' has similar preference with the `child' on movie behavior. However, on book behavior, the preferences of `woman' and `men' are more similar. Therefore, we need to utilize multiple vectors to represent users' different preference similarities.

In general, there are several challenges to solve this problem. First, although we can learn the node embedding for users on each preference separately, we will need a lot of repeated parameters and training data. How to utilize a single vector space to represent multiple aspects of similarity is a nontrivial problem. Second, social users often have relevant preferences on multiple behaviors, it is necessary to model complex interactions among multiple users' behaviors and jointly learn these in a united framework. To address the challenges mentioned above, in this paper, we propose an end-to-end framework named MCNE to learn multiple conditional node embeddings of social users. Specifically, we design a binary mask layer to separate the single node vector into multiple conditional vector subspaces corresponding to each users' preference. Then we introduce the attention network to calculate weights of the users on different preferences. And according to these learned weights, we accumulate each preference similarity message to obtain the vector representation of multi-aspect preference similarities between users, and update the node embedding by iteratively passing and aggregating the information in immediate neighborhood. Finally, we utilize the Bayesian Personalized Ranking (BPR) loss function to learn the users' preference similarity on each behavior, and jointly learn the multiple conditional node representations though multi-task learning framework. Generally, the technical contributions of our paper could be briefly summarized as follow:
\begin{itemize}
	\item we propose a novel problem that we learn multiple conditional node representations to represent multiple aspects of similarity between nodes within a single vector space.
	
	\item We design a binary mask layer to divide the single vector as multiple conditional embeddings, and introduce the attention network to model the complex interactions among multi-aspect similarities. And we utilize the multi-task framework to jointly learn conditional node embeddings.
	
	\item Extensive experiments on public datasets validate that our MCNE framework could outperform several state-of-the-art baselines with significant margin. Besides, we demonstrate that MCNE could well support the visualization and transfer learning tasks with excellent interpretability and robustness.
	
\end{itemize}

%
%

\section{Related work}\label{relatedwork}
\subsection{Network Representation Learning}
In recent years, unsupervised network representation learning methods that only utilized the network structure information are the most studied in this field. These approaches can be divided into three categories: The first is based on truncated random walks and assumes that nodes with the similar network structure have similar vector representation. DeepWalk \cite{perozzi2014deepwalk} first attempts to generate training samples by random walk on the network, and utilizes skip-gram model proposed in Word2vec \cite{mikolov2013efficient} to learn the vector representation of nodes. Noticing that DeepWalk uses the uniform sampling to generate the training sentences, node2vec \cite{grover2016node2vec} conducts the weighted random walk by two hyperparameters $p$ and $q$, in order to capture the homogeneity and structure equivalence respectively. The second is based on $k$-order distance between nodes in network. For example, LINE \cite{tang2015line} focuses on preserving first-order proximity and second-order proximity to learn the node representation. Then GraRep \cite{cao2015grarep} further captures $k$-order relational structure information to enhance node representation by manipulating global transition matrices. The third is based on deep learning techniques. With the advantage of deep learning, we can obtain higher-order nonlinear representation. Therefore, SDNE \cite{wang2016structural} proposes a semi-supervised auto-encoder model to obtain node embedding by preserving the global and local network structure information. DNGR \cite{cao2016deep} adopts a random surfing model to capture the graph structural information and learns the node representation from PPMI matrix by utilizing stacked denoising auto-encoder. GraphGAN \cite{wang2017graphgan} proposes an innovative graph representation learning framework that the generator learns the underlying connectivity distribution and the discriminator predicts the probability the edge existence between a pair of vertices. GraphSAGE \cite{hamilton2017inductive} iteratively generates the node embedding by sampling and aggregating features from the nodes' local neighborhood. And GAT \cite{velivckovic2017graph} leverages the self-attentional layers to replace the graph convolution operation. 



Furthermore, some research work formalize it into a supervised problem to obtain the task-specific node embedding. TriDNR \cite{pan2016tri} learns node representation by modeling the inter-node relationship, node-word correlation and label-word correspondence simultaneously. LANE \cite{huang2017label} proposes to learn the representations of nodes, attributes, labels via spectral techniques respectively, and projects them into a common vector space to obtain the node embedding. M-NMF \cite{wang2017community} utilizes a novel Modularized Nonnegative Matrix Factorization to incorporate the community structure into network embedding. GCN \cite{kipf2016semi} is based on an efficient variant of convolution neural network which operates directly on graphs and optimizes the node representation in semi-supervised learning graph framework. And PinSage \cite{ying2018graph} designs effective graph convolutional architectures to learn the similar relationship of graph-structured items for web-scale recommendation. And more related work on network embedding can be found in this survey \cite{cui2018survey}. Different from previous work, we propose a novel problem that learns multiple conditional network representations to represent the multi-aspect similarities of nodes in different semantics.


\subsection{Binary Neural Network}
Recently, several approaches have been proposed on the development of neural networks with binary weights \cite{courbariaux2015binaryconnect,hubara2016binarized}. The main goal of these methods is to simplify the calculations in neural networks and reduce the size of model storage. Matthieu et al. \cite{courbariaux2015binaryconnect} propose to binarize the weights for all layers during the forward and backward propagations while keeping the real-valued weights during the parameter update. The real-valued updates are found to be necessary for the application of Stochastic Gradient Descent (SGD) algorithms. Mohammad et al. \cite{rastegari2016xnor} introduce a weight binarization scheme where both a binary filter and a scaling factor are estimated. Motivated by these work, we propose a binary mask layer to automatically select the relevant embedding dimension for different tasks. To the best of our knowledge, our proposed MCNE model is the first to introduce the binarization technique in the field of network representation learning.

\begin{figure*}[t]
	\centering
	\includegraphics[width=1.0\linewidth]{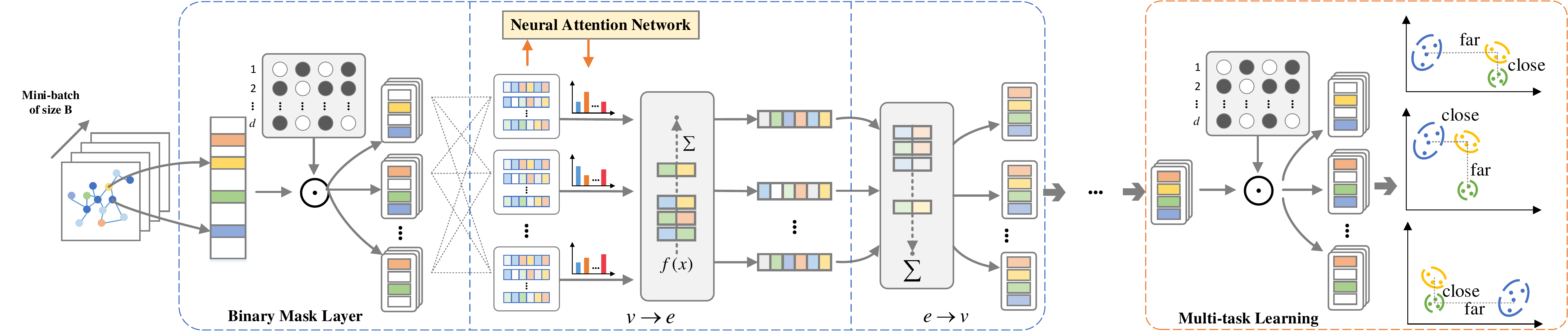}
	\vspace{-6mm}
	\caption{Framework of Multiple Conditional Network Embeddings (MCNE).}\label{fig:framework}\vspace{-0mm}
	
\end{figure*}

\section{Problem Definition}\label{preliminaries}
In this section, we will give some formal definitions of the problem for a better explanation. We first define a social network as follows:

\newtheorem{problem}{Definition}
\begin{problem}\label{def_problem_1}
	{\textbf{(Social Network)} A social network is denoted as $G=(V, E)$, where $V=\{v_{1},...,v_{n}\}$ represents the set of vertices and $E=\{e_{i,j}\}_{i,j=1}^{n}$ is the set of edges between vertices. Each edge $e_{i,j} \in E$ is associated with a weight $w_{ij}\ge 0$, which indicated the strength of relationship between vertex $i$ and vertex $j$.}
\end{problem}
In real social networks such as Facebook and Epinions, vertices often represent users in the network, and edges denote the friend or trust relationship between users. The weights on the edges are often represented by binary values that means $w_{ij}=1$ indicates user $v_{i}$ and $v_{j}$ are linked by an edge, and vice versa. 

And with the rapid development of social network, the services provided to social users have become more diverse. The users can not only establish friend relationship with each other, but also consume on different types of social network services like movies, music, etc. Therefore, the users also generate many different categories of behavioral record information in the social network, and we formally define these as follows:
\newtheorem{problem2}[problem]{Definition}
\begin{problem2}\label{def_problem_2} 
	{\textbf{(Multi-category User Behavior)} Given the social users $V\, (|V|=N)$ and items $I_{c}\,(|I_{c}|=M_{c})$ of category $c$, we utilize matrix $R_{c}\in \mathbb{R}^{|N|\times |M_{c}|}$ to represent the users' behavior record information on social service of category $c$. If user $i$ consumes item $j$, the corresponding value $R_{ij|c}=1$, otherwise it equals 0. And we utilize a set of matrices $S_{R}=\{R_{1},...,R_{C}\}$ to denote all behavior record information of social users on multiple category social services, where $C$ is the number of categories.}
\end{problem2}
As shown in the definition above, each behavior record matrix $R_{c}$ reflects the social users' preference on category $c$. However, as we illustrated the toy example in Introduction Section, the similarities between users' preferences on different categories are not identical. Therefore, it is inappropriate to only learn a single vector representation for each user to represent multiple similarity relationships between users. In order to address this problem, we first elaborate the formal definition as follows: 
\newtheorem{problem3}[problem]{Definition}
\begin{problem3}\label{def_problem_3}
	{\textbf{(Multiple Conditional Network Representations)} Given a network $G=(V, E)$ and a set of multi-category behavior matrices $S_{R}=\{R_{1},...,R_{C}\}$, we aim to simultaneously learn a set of low-dimensional conditional vector representations $S_{U}=\{U_{1},...,U_{C}\}$ for social users on multiple category behaviors. And each conditional vector representation $U_{c}\in \mathbb{R}^{|V|\times d}(d<<|V|)$ should satisfy the following properties: 1) the conditional network representation should preserve the network structure information; 2) the conditional network representation should maintain the similarity relationship of users' behavior on category $c$.}
\end{problem3}
Next we will introduce how our proposed model can simultaneously learn multiple conditional network representations in a united vector space.

\section{MCNE: Multiple Conditional Network Embeddings}\label{model}
In this section, we first present a general description of our model. Then we introduce each part of the model in detail, and finally illustrate the model optimization.  

\subsection{Framework}
In this paper, we propose the \textbf{M}ultiple \textbf{C}onditional \textbf{N}etwork \textbf{E}mbedding (MCNE) model to jointly learn the network structure and multi-category user behavior information, which is illustrated in Figure ~\ref{fig:framework}. Specifically, we adopt the framework of Graph Neural Network (GNN) that is based on the message-passing-receiving mechanism, in order to iteratively aggregate information from a node's local neighborhood and update the node representation. For each layer of graph neural network, we first utilize a binary mask layer to select the relevant vector dimensions corresponding to each user's behavior preference. Then we use the attention mechanism to calculate the weights of different behaviors between adjacent users, and we aggregate the multi-preference information according these weights to update the node representation of next layer. Furthermore, we utilize Bayesian Personalized Ranking (BPR) loss function to learn the users' preference similarity on each behavior. Finally, we use the multi-task framework to simultaneously learn multiple conditional network representations, in order to represent different preference similarities of social users. As shown in Figure ~\ref{fig:framework}, MCNE mainly contains three parts, i.e., generating multiple conditional network representations, learning users' preference similarity on a specific behavior, and jointly learning multiple user preferences. Next we will elaborate the technical details of each part.

\subsection{Generating Multiple Conditional Network Representations}\label{generating_embedding}
In this section, we describe how to generate multiple conditional network representations for each node. 

\subsubsection{Embedding Layer}
Similar as many graph neural network models, the input of our model is a social network $G(V, E)$. We first project all users into a low-dimensional vector space and utilize the matrix $U^{0}\in \mathbb{R}^{|V|\times d_{0}}$ to represent the initial node representation. We regard these node vectors as a comprehensive representation that can denotes users' preferences on all behaviors. In practical applications, we can also process the users' profiles, such as gender and age, as the initial node representation, and formalize it into an inductive representation learning problem. While it's not the focus of this paper, we leave the exploration as a future work.   

\subsubsection{Binary Mask Layer}
From the previous part, we obtain the initial node representation $U^{0}$, which can also be considered as the node embedding of the $k$-th ($k$=0) layer of the graph neural network. In order to model the multiple conditional similarities betweens users in different behaviors, we introduce a binary mask layer to divide the node representation of each layer into different vector subspaces, which is illustrated in the leftmost part of Figure ~\ref{fig:framework}. Specifically, our modified graph neural network associates the node embedding of each layer with a real-valued mask weight matrix $M^{k}_{r}\in \mathbb{R}^{|C+1|\times d_{k}}$, where $d_{k}$ is the dimension size of node embedding in $k$-th layer and $C$ is the number of user behavior category. And we add an additional dimension to represent other user behavior preferences that are not included in the training dataset. After that, we obtain the binary mask matrix $M^{k}_{b}$ by passing the real-value mask matrix $M^{k}_{r}$ through a hard thresholding function given by 
\begin{equation}\label{1}
M^{k}_{b_{ij}}=
\begin{cases}
1, &  \text{if}\  M^{k}_{r_{ij}}\ge 0 \\
0, & \text{otherwise}
\end{cases},
\end{equation}
we denote $m_{c}^{k}$ to be the selection of the $c$-row of binary mask matrix $M^{k}_{b}$. And the conditional network representation of user $i$ on $c$-th behavior can be defined as follows:
\begin{equation}\label{2}
u_{i\shortmid c}^{k}=u_{i}^{k}\odot m_{c}^{k},
\end{equation}
where $u_{i}^{k}$ is the node embedding of user $i$ in $k$-th layer, and $\odot$ represents the element-wise product of two vectors. And the mask $m_{c}^{k}$ plays the role of a gating function selecting the dimension of embedding related to $c$-th behavior depending on whether the corresponding value is 0 or 1. By using the binary mask matrix $M^{k}_{b}$, we can divide a united vector space of node embedding into different vector subspaces related to each behavior. For example, a dense node embedding such as [0.1,0.9,-0.5,1] can obtain multiple conditional network representations such as [0.1,0,0,1], [0,0.9,-0.5,0] and [0,0.9,-0.5,1] after different binary masks. And we consider these separate subspaces as the conditional network representations of the user on different behaviors. Although the hard thresholding function in Eq.~\eqref{1} is non-differentiable, we combine the techniques used in network binarization \cite{courbariaux2015binaryconnect} and train these mask variables in end-to-end fashion, as described in detail below.  

\subsubsection{Multi-aspect Similarity Message Sending Operation}
\begin{figure}[t]
	\centering	
	\includegraphics[width=0.45\textwidth]{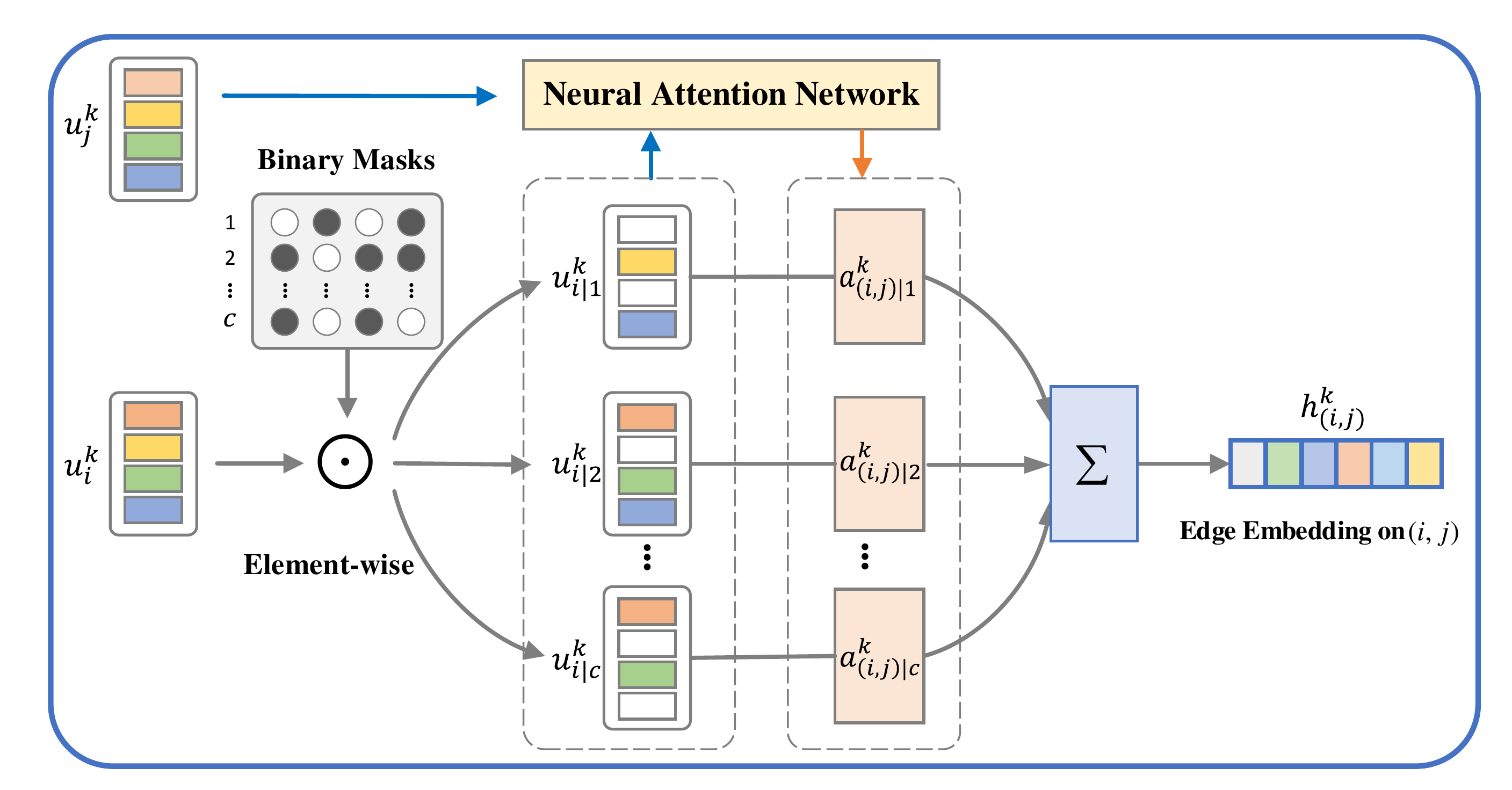}\vspace{-1mm}
	\caption{Multi-aspect Similarity Message Sending Operation in MCNE.}\vspace{-4mm}
	\label{model:fig3}
\end{figure}

From the binary mask layer, we can obtain the multiple conditional node representations in $k$-th layer. Then we adopt the message sending and receiving operations \cite{scarselli2009graph} of graph neural network to get the node embedding in next ($k$+1)-th layer. In this part, we first introduce our modified multi-aspect similarity message sending operation, which is illustrated in Figure ~\ref{model:fig3}. Specifically, the multi-aspect similarity message $h_{(i,j)}$ sent by node $v_{i}$ on the connected edge $e_{i,j}$ is defined as follows:
\begin{equation}\label{3}
v\to e: h_{(i,j)}^{k}=\sum_{c=1}^{|C|+1}a_{(i,j)|c}^{k}\cdot u_{i|c}^{k},
\end{equation}
where $u_{i|c}^{k}$ is the conditional representation of users on $c$-th behavior, and $a_{(i,j)|c}^{k}$ is the corresponding weight. Then we accumulate all the conditional representation according to the weights, and obtain the embedding $h_{(i,j)}^{k}$ of edge $e_{i,j}$ that contains multiple users' preference information. As we illustrated in Figure ~\ref{introduction:fig1}, the preference similarities of users on different behaviors are not equivalent. So we introduce the attention network to calculate weight score $a_{(i,j)|c}^{k}$, as described below:
\begin{equation}\label{4}
a_{(i,j)|c}^{k'}=\textbf{h}^{k^{T}}ReLU(\textbf{W}^{k}_{a}[u_{i|c}^{k},u_{j|c}^{k}]) ,
\end{equation}
where $\textbf{W}^{k}_{a}\in \mathbb{R}^{t\times 2d^{k}}$ and $\textbf{h}^{k}\in \mathbb{R}^{t}$ are model parameters. We take each conditional representation $[u_{i|c}^{k},u_{j|c}^{k}]$ of two adjacent nodes on edge $e_{i,j}$ as input of the attention network, and then obtain the corresponding weight score by a multi-layer neural network. Furthermore, these scores are normalized by the softmax function:
\begin{equation}\label{5}
a_{(i,j)|c}^{k} = \frac{{\rm exp}\{a_{(i,j)| c}^{k'}\}}{\sum_{l=1}^{|C|+1}{\rm exp}\{a_{(i,j)|l}^{k'}\}} , 
\end{equation}
by using this attention network, more similar behavior between users will be assigned a greater weight score. And the greater the weight score $a_{(i,j)|c}^{k}$ of $c$-th behavior corresponds, the more message of conditional representation $u_{i|c}^{k}$ will be passed on edge $e_{i,j}$, which makes the users more similar on $c$-th behavior.  

\subsubsection{Multi-aspect Similarity Message Receiving Operation}
After we obtain the representation of message passed on edges, we further take the message receiving operation to update the node embedding in next layer. For each node $v_{i}$ in the network, the detailed process of receiving operation is illustrated as follows:
\begin{equation}\label{6}
e\to v: 
\begin{cases}
\quad h_{\mathcal{N}(i)}^{k+1}=\text{AGGR}^{k+1}(\{h_{(i,j)}^{k}, \forall j\in \mathcal{N}(i)\})\\
\quad u_{i}^{k+1}=\text{ACT}(\textbf{W}^{k+1}[h_{\mathcal{N}(i)}^{k+1},u_{i}^{k}])
\end{cases},
\end{equation}
\vspace{1mm}
specifically, we uniformly sample a fixed-size set of nodes $\mathcal{N}(i)$ from the neighbors of $v_{i}$\footnote{Exploring weight sampling is an important direction for further work.}, in order to keep the computational footprint of each batch fixed. Then we utilize the average pooling as the function $\text{AGGR}^{k+1}$ to aggregate the neighbors' multi-aspect similarity messages into a single vector $h_{\mathcal{N}(i)}^{k+1}$. Finally, we combine the current node embedding $u_{i}^{k}$ and aggregated neighborhood vector $h_{\mathcal{N}(i)}^{k+1}$, and take them into a fully connected layer with nonlinear activation function $\text{ACT}(x)=max(0,x)$ to obtain the node embedding $u_{i}^{k+1}$ in next ($k$+1)-th layer. 

\subsubsection{Final Multiple Conditional Network Representations}
By the multi-aspect similarity message sending and receiving operation in each layer, we are able to aggregate the information of multiple users' preferences in immediate neighborhood. Then we can continuously stack $k$ layers to capture more information from $k$-order nodes. And we denote the final node embedding output at last layer ($k$+1), as $U\in \mathbb{R}^{|V|\times d}$. Furthermore, we utilize the last binary mask layer to obtain final multiple conditional network representations $\{U_{1}, U_{2},...,U_{C}\}$ of social users on different behaviors.

\subsection{Jointly Learning of Conditional Network Representations}
In this section, we will introduce how to simultaneously learn multiple conditional network representations of social users on different behaviors. As we focus on the implicit feedbacks of users, we utilize the widely used Bayesian Personalized Ranking (BPR) criterion to learn the users' preference similarity in each behavior. Given the record matrix $R_{c}$ and conditional embedding $u_{i|c}$ on $c$-th behavior, the loss function is formally defined as follows: 
\begin{equation}\label{7}
\mathcal{L}_{c}(R_{c}, U_{c})= -\sum_{(i,p,n)}{\rm ln}\, \sigma(u_{i|c}z_{p|c}-u_{i|c}z_{n|c})+\lambda_{1}\|\Theta_{c} \|^{2},
\end{equation}
where $z_{p|c}$ and $z_{n|c}$ are the embeddings of item $p$ and $n$ on $c$-th behavior, with the same dimension as the conditional node embedding $u_{i|c}$. For each user $v_{i}$, we take the interacted item $p$ as the positive sample, and randomly select the non-interacted item $n$ as the negative sample. By optimizing Eq.~\eqref{7}, we can make the users' preference on positive item larger than those on negative item, so that we can learn the users' preference similarity on $c$-th behavior.  

Meanwhile, the multiple behaviors of users are often relevant in practical application. To this end, we consider the learning of each conditional embedding as a single task, and utilize the multi-task learning framework to jointly learn the multiple conditional embeddings, as defined below:  
\begin{equation}\label{8}
\mathcal{L}(S_{R},S_{U})=\frac{1}{C}\sum_{c=1}^{C}\mathcal{L}_{c}(R_{c}, U_{c})+\lambda\|\Theta\|^{2},
\end{equation}
where $C$ is the number of user behaviors and tasks, $\Theta$ indicates all the parameters of our model. Through the combination of our binary mask layer and multi-task learning framework, we can simultaneously learn users' multiple preference similarities on different behaviors, and utilize the learnable subspace to share the relevant information between different behaviors, which can effectively alleviate the sparsity of training data, and enhance the robustness of learned multiple conditional network representations. 

\subsection{Model Optimization}
In this section, we will introduce the optimization of our model in detail, which consists of two parts: binary mask learning and mini-batch training algorithms.
\subsubsection{Binary Mask Learning} Because the initialization of real-valued mask weight matrix $M_{r}^{k}$ has a great impact on the convergence and performance of our model, we will describe it in detail here. We utilize the Uniform Distribution $M^{k}_{r_{ij}}\sim U(-0.5,0.5)$ to initialize the variables, and find that initialization with a fixed value can not achieve competitive performance in experiments. And this initialization scheme proposed above allows us to better understand the benefits of learnable masks, which we will further elaborate in the following Experiments Section.  

In Eq.~\eqref{1}, we obtain the binary mask matrix $M_{b}^{k}$ by applying a non-differentiable threshold function to the real-valued matrix
$M_{r}^{k}$. In order to use the class of Stochastic Gradient Descent (SGD) algorithms to update the variables in matrix $M_{r}^{k}$, we adopt the training method proposed in \cite{courbariaux2015binaryconnect}. Specifically, we only binarize the mask variables during the forward and backward propagations of network, and update the real-valued mask variables $M_{r}^{k}$ using the gradients computed for binary mask variables $M_{b}^{k}$. In addition, we enforce the real-valued variables to lie within the $\left[-1,1\right]$ interval at each training iteration. Because the real-valued variables would otherwise grow very large with any impact on the binary variables. Finally, we utilize the Adam algorithm\cite{kingma2014adam} to train our model in an end-to-end differentiable manner.

\subsubsection{Mini-batch Training} In Section \ref{generating_embedding}, the computational complexity of generating multiple conditional network representations for all nodes is very high. So we extend our proposed generating algorithm to the mini-batch setting, in order to enable that our model can be applied to the large-scale social networks. Concretely, given a part of nodes in the network, we first forward sample the required neighbors (up to layer $k$+1), and only compute the node representations that are necessary to used in each layer, which can greatly improve the speed of our model optimization.

\textbf{Time Complexity.} By using the mini-batch training setting, the time complexity for our MCNE model is fixed at $\mathcal{O}(M\prod_{i=1}^{k}|\mathcal{N}_{i}|)$, where $M$ is the number of nodes in each mini-batch data, $k$ is the layer number of our modified graph neural network, and $\mathcal{N}_{i}$ is the number of sampling neighbors in each layer. In general, we often set $k=2$ and $|\mathcal{N}_{1}|\cdot |\mathcal{N}_{2}|\le 50$ to achieve satisfactory performance. Therefore, the time complexity is acceptable and the proposed MCNE model could be applied to the real-world applications. 

\section{Experiments}\label{experiments}

\subsection{Experimental Settings}\label{sec:experimental_result}
\subsubsection{Datasets} 
We conduct the experiments on two publicly available datasets Ciao and Epinions\footnote{https://www.cse.msu.edu/\textasciitilde tangjili/trust.html}, which are two popular who-trust-whom online social networks. The users consume multiple different categories of products on the website, and establish their trust networks based on other users' product views. And we select three representative behaviors from Ciao and Epinions to study multiple preferences of users in the network, respectively. 

In the dataset preprocessing step, for both datasets, we filtered out the users that have less than 2 social links and 2 multi-category behavior records. The detailed statistics of the data after preprocessing are shown in Table ~\ref{tab:datasets}.

\begin{table}
	\centering
	\caption{Statistics of the Datasets}\vspace{-3mm}
	\label{tab:datasets}
	\resizebox{0.48\textwidth}{15mm}{
		\begin{tabular}{c|c|c|c|c|c}
			\hline	
			\hline
			\textbf{Datasets} & \textbf{\#Users} & \textbf{\#Links} & \textbf{Behaviors} & \textbf{\#Items} & \textbf{\#Interactions}\\
			\hline
			\hline
			\multirow{3}{*}{Ciao} & \multirow{3}{*}{4,321}	 &  \multirow{3}{*}{121,408} & Beauty & 9,243 & 23,091\\ 	\cline{4-6}
			& & & Book & 12,409 & 21,105\\	\cline{4-6}
			& & & Travel & 11,899 & 20,857\\	\cline{4-6}
			\hline
			\hline
			\multirow{3}{*}{Epinions} & \multirow{3}{*}{10,459} &  \multirow{3}{*}{280,258} & Game & 6,804 & 30,417\\ 	\cline{4-6}
			& & & Electronics & 12,425 & 30,429\\	\cline{4-6}
			& & & Travel & 11,885 & 38,578\\	\cline{4-6}
			\hline
			\hline
	\end{tabular}}\vspace{-0.5mm}
\end{table}

\begin{table*}
	\centering
	\caption{Recall@5 and NDCG@5 Comparisons for Multiple Behaviors on the Ciao and Epinions Datasets.}\vspace{-3mm}
	\label{tab:result_cmp}
	\resizebox{\textwidth}{21mm}{
		\begin{tabular}{c||c|c|c|c|c|c||c|c|c|c|c|c}
			\hline	
			\hline
			Datasets & \multicolumn{6}{|c||}{Ciao} & \multicolumn{6}{|c}{Epinions}\\
			\hline
			Tasks & \multicolumn{2}{|c}{Beauty} & \multicolumn{2}{|c}{Book} &  \multicolumn{2}{|c||}{Travel} &  \multicolumn{2}{|c}{Electronics} &  \multicolumn{2}{|c}{Travel} &  \multicolumn{2}{|c}{Game} \\
			\hline
			Metrics & Recall@5  & NDCG@5  & Recall@5  & NDCG@5  & Recall@5  & NDCG@5   & Recall@5 & NDCG@5   & Recall@5  & NDCG@5  	& Recall@5  & NDCG@5  \\
			\hline
			BPR &	12.71\% &	21.14\% &	14.75\% &	21.23\% &	14.73\% &	16.20\% &	11.83\% &	13.64\% &	30.96\% &	33.29\% &	34.47\% &	39.86\%\\
			LibFM &	15.38\% &	22.34\% &	16.42\% &	21.62\% &	15.31\% &	20.33\% &	18.93\% &	18.78\% &	32.73\% &	36.92\% &	35.62\% &	40.97\% \\
			NFM	 & 16.79\% &	23.94\% &	18.24\% &	22.51\% &	15.91\% &	21.36\% &	19.37\% &	19.61\% &	33.83\% &	37.36\%  &	36.57\% &	41.55\%\\
			\hline
			Node2vec-B &	19.17\% &	26.88\% &	21.98\% &	27.36\% &	19.78\% &	24.68\%  &	24.33\% &	23.87\% &	39.24\% &	42.07\% &	41.14\% &	45.64\%\\
			LINE-B &	19.49\% &	28.80\% &	19.71\% &	26.50\% &	16.21\% &	21.18\% &	23.35\% &	24.41\% &	38.03\% &	41.19\%
			& 39.51\% & 44.07\%\\
			GraphSage	& 17.25\% &	25.62\% &	20.31\% &	25.31\% &	18.32\% &	23.84\% &	25.12\% &	23.11\% &	36.12\% &	40.85\%
			&	40.12\% &	43.47\%\\
			PinSage &	23.41\% &	31.01\% &	25.13\% &	28.55\% &	24.45\% &	29.47\% &	27.36\% &	26.52\% &	42.56\% &	44.02\%
			&	42.68\% &	45.54\%\\
			\hline
			MCNE &	\textbf{28.18\%} &	\textbf{36.40\%} &	\textbf{29.67\%} &	\textbf{35.75\%} &	\textbf{28.47\%} &	\textbf{34.75\%} &\textbf{33.20\%} &	\textbf{33.84\%} & \textbf{45.30\%} &	\textbf{48.24\%} & \textbf{46.98\%} &	\textbf{50.92\%}\\
			\hline
			\hline
	\end{tabular}}
\end{table*}

\subsubsection{Baselines}
In order to demonstrate the effectiveness of learned conditional network representations, we compare with the-state-of-art algorithms and some variants of MCNE. And the compared baselines are selected from two aspects. One is the representative algorithms in recommendation field like BPR, LibFM and NFM, and the other is competitive network representation learning methods such as Node2vec, LINE, GraphSage and PinSage. And the details of these baselines are illustrated as follows:
\begin{itemize}
	\item \textbf{BPR}\cite{rendle2009bpr}: It's a competitive latent factor model for implicit feedback based recommendation, which designs a ranking based function that assumes user prefers the interacted items over all other non-observed items.
	
	\item \textbf{LibFM}\cite{rendle2012factorization}: It's the official implementation\footnote{http://www.libfm.org/} of Factorization Machines. And it has shown strong performance for personalized tag recommendation and context-aware prediction\cite{rendle2011fast}. We utilize the user's adjacency matrix as its own attributes.
	
	\item \textbf{NFM}\cite{he2017neural}: It utilizes the deep neural network to capture the higher-order feature interaction, and bring together of linear factorization machines with the strong representation ability of non-linear neural networks.
	
	\item \textbf{Node2vec}\cite{grover2016node2vec}: Different from DeepWalk, it designs a biased truncated random walks to efficiently explore diverse neighborhood and utilize skip-gram model to learn node embedding. Besides, we combine Node2vec and BPR to formalize a supervised network representation learning problem, and obtain node embedding by jointly predicting the network structure and items consumed by users. We represent this supervised algorithm as \textbf{Node2vec-B}, and only report the best results of them for brevity.
	
	\item \textbf{LINE}\cite{tang2015line}: It learns the network embedding by preserving the first-order proximity or second-order proximity of the network structure separately. And we utilize the best of them as the final baseline. Similar to Node2vec, we also add the \textbf{LINE-B} method and only illustrate the best algorithm in the experimental results.
	
	\item \textbf{GraphSage}\cite{hamilton2017inductive}: It iteratively generates the node embedding by sampling and aggregating features from the nodes' local neighborhood. Then the node embedding are learned by maintaining the network structure information.
	
	\item \textbf{PinSage}\cite{ying2018graph}: It's a state-of-the-art algorithm designing effective graph convolutional architectures for web-scale recommendation. The origin PinSage focuses on learning the embeddings of graph-structured items, and we generalize their method to apply to our problem by constructing a complete network by user-user and user-item bipartite graphs. 
	
	\item \textbf{MCNE-F}: It is the simplified version of MCNE, which utilizes the fixed disjoint masks instead of our proposed learnable binary masks, illustrated in the left of Figure~\ref{fig:mask}, and omits the attention network. 
	
	\item \textbf{MCNE-A}: It is the reduced version of MCNE only without introducing the attention network.
\end{itemize} 

\subsubsection{Evaluation Protocols} 
In order to evaluate the performance of learned conditional network representations, we randomly select 70\% of each behavior data as training set, 10\% as validation set, and 20\% as test set. For the compared baselines, we adopt two different dataset partitioning methods. One is that we conduct the baselines on the training sets of all behavior records, and regard the learned single vector representation as conditional embeddings of users on all behaviors. Another is that we independently train these baselines on the training set of each behavior, so as to obtain the different conditional representations of users. And we take the best of them as the final experimental results for brevity. Meanwhile, we utilize BPR model on the training set to learn the item embeddings for unsupervised network representation learning methods like Node2vec, LINE and GraphSage. And for the supervised algorithms, we can directly use the inner product of the conditional node embeddings and item embeddings as the similar score. Since it's too time-consuming to rank all items for every user during evaluation, we adopt the common strategy\cite{yang2011like,wu2017modeling} that randomly sample 100 negative items that are not interacted by the user, and rank the test items among the 100 items. Furthermore, the performance of a ranked list is evaluated by two widely adopted ranking based metrics: Recall and Normalized Discounted Cumulative Gain (NDCG). And we truncate the ranked list at different value $K$=$[5, 10, 20]$ for both metrics, and observe the similar trend. So we only report the results of $K$=5 for brevity. Finally, we repeat each experiment 10 times independently and report the average ranking results. 

\subsubsection{Parameter Settings}
We implement our method based on TensorFlow, and the model parameters are initialized with a Gaussian distribution(with a mean of 0 and standard deviation of 0.01). We set the number of hidden layers $k$ in MCNE as 2, the embedding dimensions of each layer as $[256, 128, 100]$, and neighbors sample size $\mathcal{N}_{i}$ of each layer as $[20, 20]$. The regularization parameter $\lambda$ is set as 0.0001 and the number of negative item samples is set as 5. In the process of model training, we set the learning rate as 0.003 and mini-batch size as 128. The parameters of other baselines are the same as those in their origin paper and tuned to be optimal. Besides, we set the final node embedding dimension to be 100 for all methods in order to get a fair comparison.  

\begin{figure}[t]
	\centering	
	\includegraphics[width=0.5\textwidth]{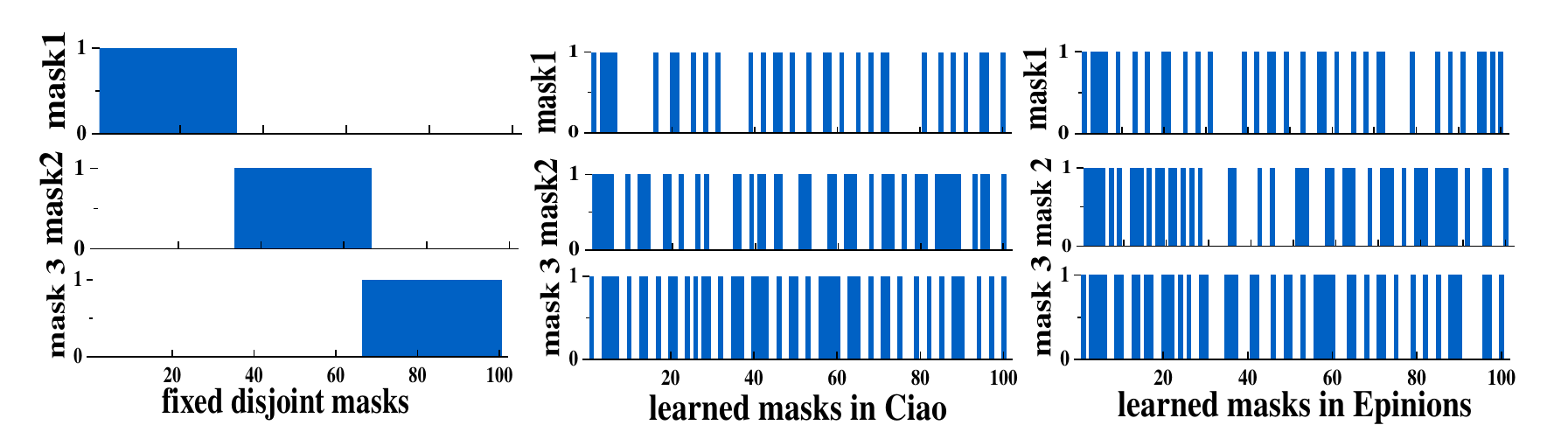}\vspace{-0.3cm}	
	\caption{Visualization of the Last Binary Masks}\label{fig:mask}\vspace{0.1cm}
	\vspace{-0.5cm}
\end{figure}

\begin{figure*}[!t]
	\begin{center}
		$\begin{array}{ccccc}
		\hspace{-0.1cm}\includegraphics[width=0.45\linewidth]{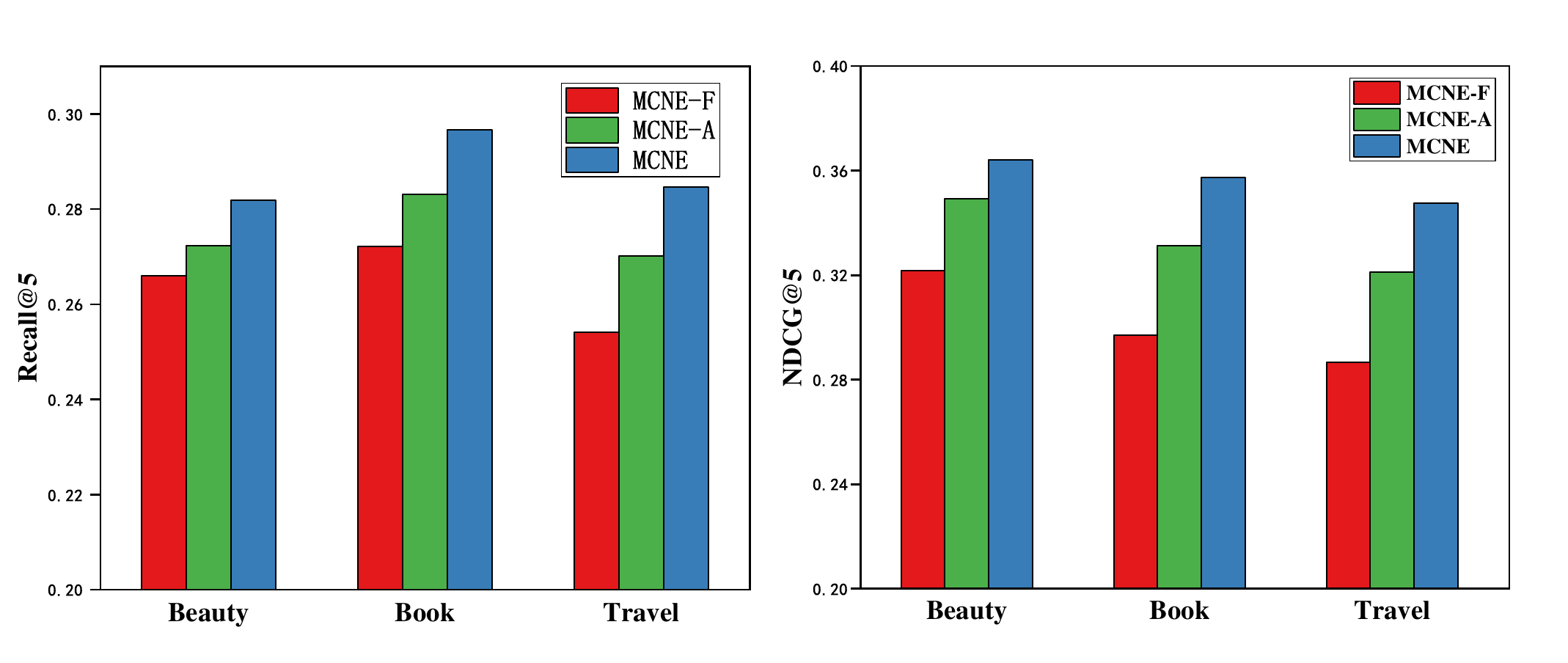} &
		\hspace{-0.1cm}\includegraphics[width=0.45\linewidth]{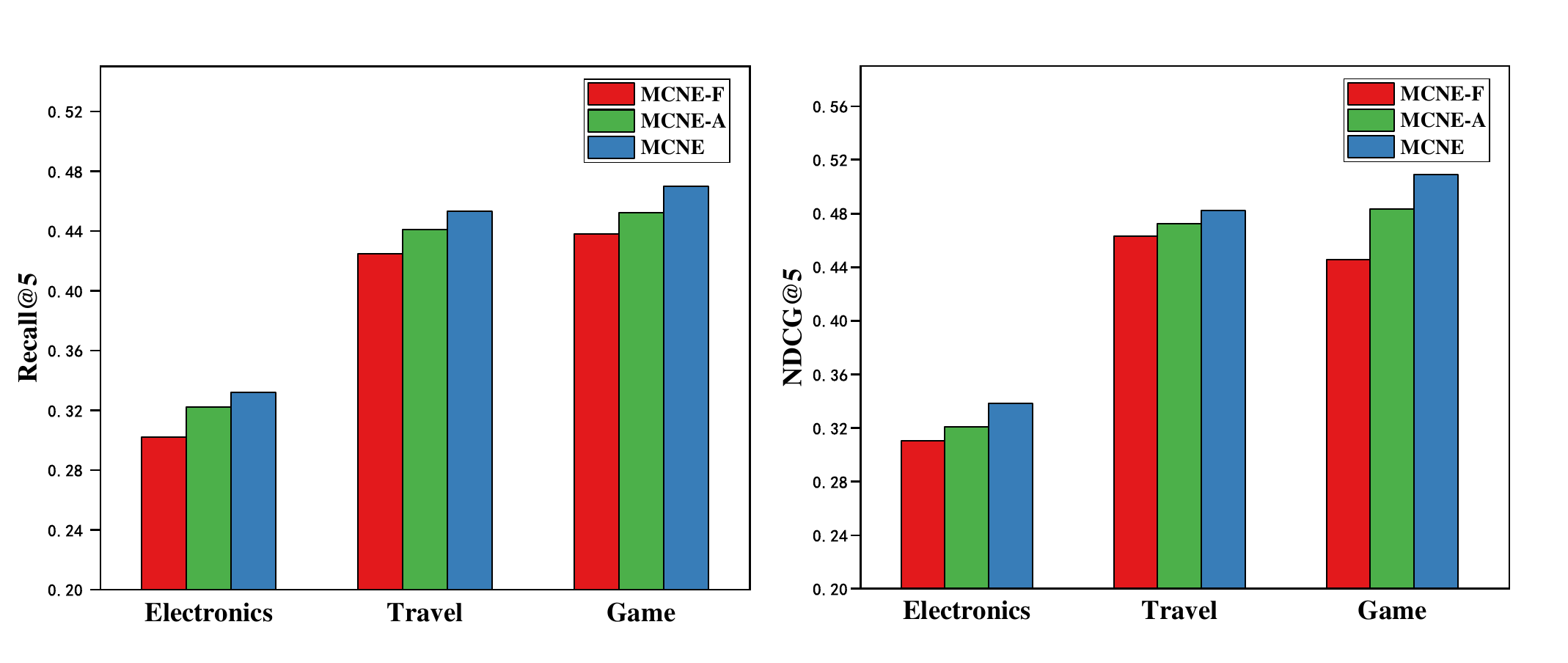}&\\
		\hspace{-0.2cm} \text{(a) Ciao} &
		\hspace{-0.2cm} \text{(b) Epinions}& \\
		\end {array}$
		\vspace{-3mm}
		\caption{\label{link_prediction}Comparison Results of MCNE Variants on Ciao and Epinions Datasets.}
		\vspace{-0mm}
	\end{center}
\end{figure*}

\subsection{Experimental Results}
In this section, we first introduce the overall performances of all models on two datasets, and then present the comparison results of MCNE variants. 
\subsubsection{Performance Comparison}
In this part, we first compare with the state-of-the-art methods on two datasets, and illustrate the Recall@5 and NDCG@5 results for both datasets in Table~\ref{tab:result_cmp}. And we utilize \textbf{bold-faced} to highlight the best experimental results. From Table~\ref{tab:result_cmp}, we can obtain the following observations:

First, compared with all the baselines, MCNE model consistently achieve significant improvements on multiple behavior tasks of two datasets. It demonstrates that the conditional network representations learned by our MCNE model can effectively represent the user's different preferences in multiple behaviors. 

Second, by comparing with the representative recommendation algorithms BPR, LibFM and NFM, we can observe that the NRL methods achieve better performance. This shows that node embedding learned by NRL methods can better capture the network structure information between users.

Third, we can find that the performance of PinSage, Node2vec-B and LINE-B outperforms GraphSage, LINE and Node2vec, respectively. It demonstrates that comprehensive node embedding learned by unsupervised NRL methods can't effectively present the preference similarity between users on a specific behavior. That means it's necessary to learn a corresponding conditional network representation for each behavior, so as to capture the different similarities. 

Finally, PinSage achieves better performance than Node2vec-B and LINE-B methods, which indicates that graph neural network framework can capture network structure information between users more effectively. And MCNE obtains better performance than PinSage, because it considers the weighting relationships between different preferences of social users, and utilizes  multi-task learning framework to share information among multiple behaviors. 

\subsubsection{Performance with MCNE Variants}
In order to demonstrate the effectiveness of each part in MCNE, we present the comparison results of MCNE-F, MCNE-A, and MCNE in Figure~\ref{link_prediction}. And we can obtain the following conclusions:

First, Compared with MCNE-F, MCNE-A achieves better performance on two datasets, which proves the effectiveness of our proposed binary mask layer. And we visualize the last binary mask $m_{c}^{k+1}$ for each behavior in Figure~\ref{fig:mask}. Then we can observe that the learned masks filters out the relevant embedding dimension for each behavior to obtain the conditional network representation, and utilize the overlapping dimensions to share relevant information.

Second, From the comparison results of MCNE-A and MCNE, we can illustrate that the proposed attention network can effectively models the weighting relationships of multiple behaviors between social users, and make the users with more similar behaviors closer in the conditional vector space. Besides, we can infer the reason for formation of edges between users according to the weighting distribution obtained by the attention network, which can enhance the interpretability of embedding learning algorithm. And we will visualize learned weight values and integrate them into model's interpretable part in further work. 

\subsection{Further Experiments}
In this section, we will introduce the conditional network embedding visualization and transfer learning for MCNE 
, in order to illustrate the interpretability and robustness of our proposed model. 

\subsubsection{Conditional Network Embedding Visualization}
\begin{figure}[t]
	\centering	
	\includegraphics[width=0.22\textwidth]{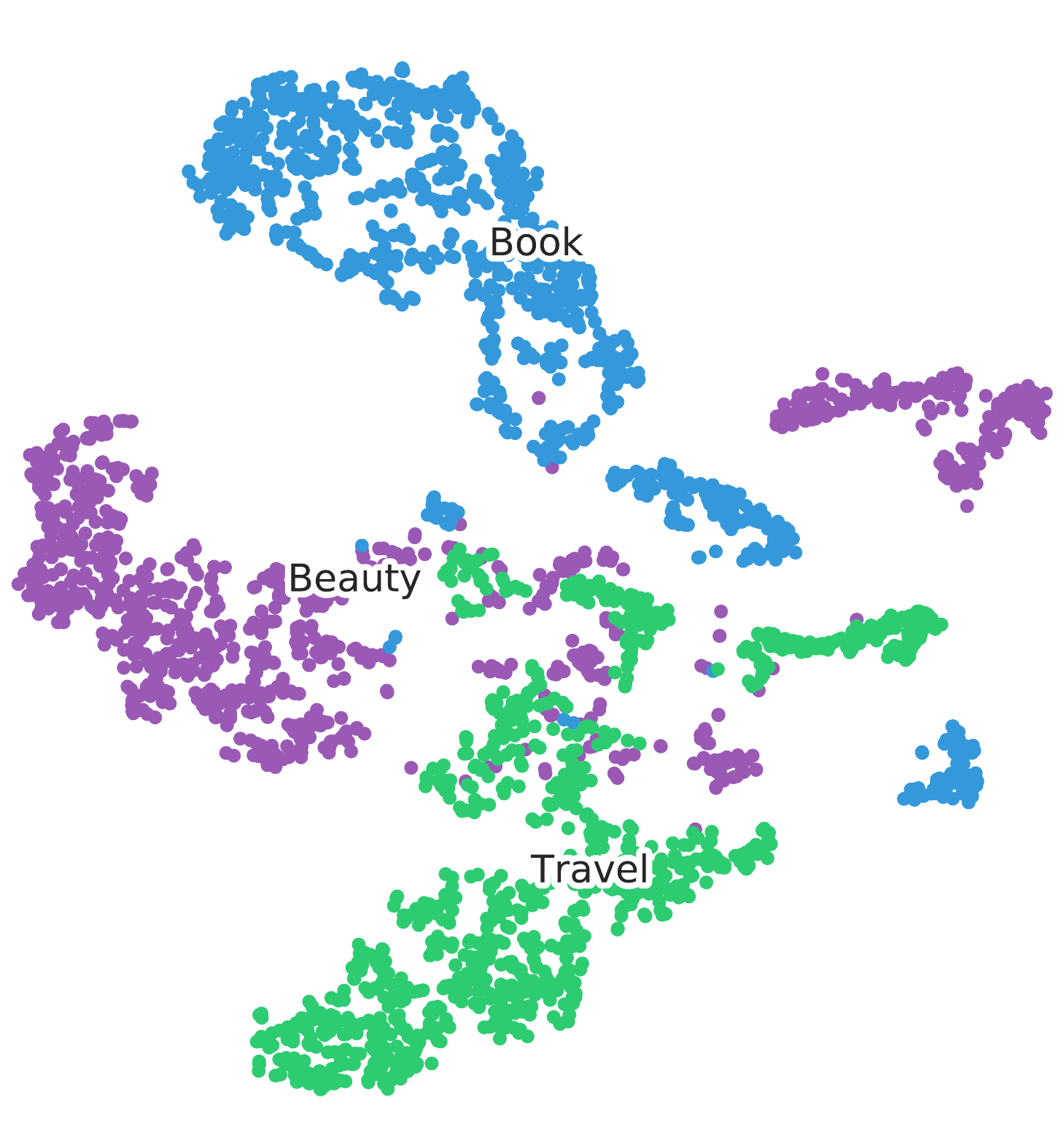}\vspace{-0.3cm}	
	\caption{Visualization of Conditional Network Representations on Ciao Dataset.}\vspace{0.1cm}
	\label{fig:embedding_vis}\vspace{-0.4cm}
\end{figure}

An important application of network embedding is to generate visualizations of a network in a two-dimensional space, so that we can intuitively observe the relevance between nodes in the network. To conduct the visualization task, we randomly select 1000 users in the Ciao dataset. Then we project their multiple conditional node embeddings learned by MCNE into a two-dimensional space by the widely used visualization tool t-SNE\cite{maaten2008visualizing}. We represent each conditional node embedding in different colors and mark the corresponding behavior name at the center of each conditional node embedding. And the visualization result is shown in Figure~\ref{fig:embedding_vis}.

From Figure~\ref{fig:embedding_vis}, we can observe that there is an obvious distance between different conditional network embeddings. It demonstrates that the multiple conditional embeddings learned by MCNE are discriminative, so they can represent different similarities of users on multiple behaviors. Meanwhile, there is a small amount of coverage between conditional network embeddings. This result can illustrate there is a certain correlation between conditional node embeddings. Through Figure~\ref{fig:mask} and Figure~\ref{fig:embedding_vis}, we can further demonstrate the effectiveness of our proposed binary mask layer and explain the correlation between conditional network embeddings.

\subsubsection{Transfer Learning for MCNE}
In this section, we want to investigate whether MCNE can transfer to a new task and achieve the satisfactory performance. In order to conduct this experiment, we add two user behavior records from the Ciao and Epinions datasets, and compare with the most competitive baseline PinSage. Specifically, we remove the attention network in MCNE and regenerate the corresponding real-valued mask variables for the new task. Then we directly use the fixed model parameters of trained MCNE, and only update the real-valued mask variables. According to the experimental setting introduced in Section~\ref{sec:experimental_result}, we finally illustrate the performance of new task in Table~\ref{tab:transfer}.

From Table~\ref{tab:transfer}, we can observe that MCNE achieves the better performance than PinSage, which proves MCNE can effectively capture multi-aspect similarities between users, and robustly transfer to the new task. Besides, different from PinSage that must retrain all the model parameters on the new task, MCNE only need to update a small number of real-valued mask variables, which can greatly improve the efficiency and applicability of our model.

\begin{table}
	\centering
	\caption{Results on Transfer Learning for MCNE}
	\label{tab:transfer}
	\resizebox{0.44\textwidth}{12mm}{
		\begin{tabular}{c||c|c||c|c}
			\hline	
			\hline
			Datasets & \multicolumn{2}{|c||}{Ciao} & \multicolumn{2}{|c}{Epinions}\\
			\hline
			Tasks & \multicolumn{2}{|c||}{Restaurant} &  \multicolumn{2}{|c}{Sports} \\
			\hline
			Metrics & Recall@5 & NDCG@5 & Recall@5 & NDCG@5 \\
			\hline
			PinSage &   28.18\% & 25.97\% & 36.41\% & 31.17\%\\
			\hline
			MCNE  & \textbf{31.20\%} & \textbf{27.55\%} & \textbf{38.03\%} & \textbf{32.69\%}\\
			\hline
			\hline
	\end{tabular}}
\end{table}

\subsection{Parameter Sensitivity}
We investigate the sensitivity of our model parameter in this section. Specifically, we mainly evaluate how the dimension of node embedding $d$ and neighbors sample size $|\mathcal{N}_{i}|$ affect the performance. We conduct this experiment on two datasets and obtain the similar trend. So we only report the results on the Ciao dataset for brevity.

\begin{figure}[t]
	\begin{center}
		$\begin{array}{ccccc}
		\hspace{-0.2cm}\includegraphics[width=0.5\linewidth]{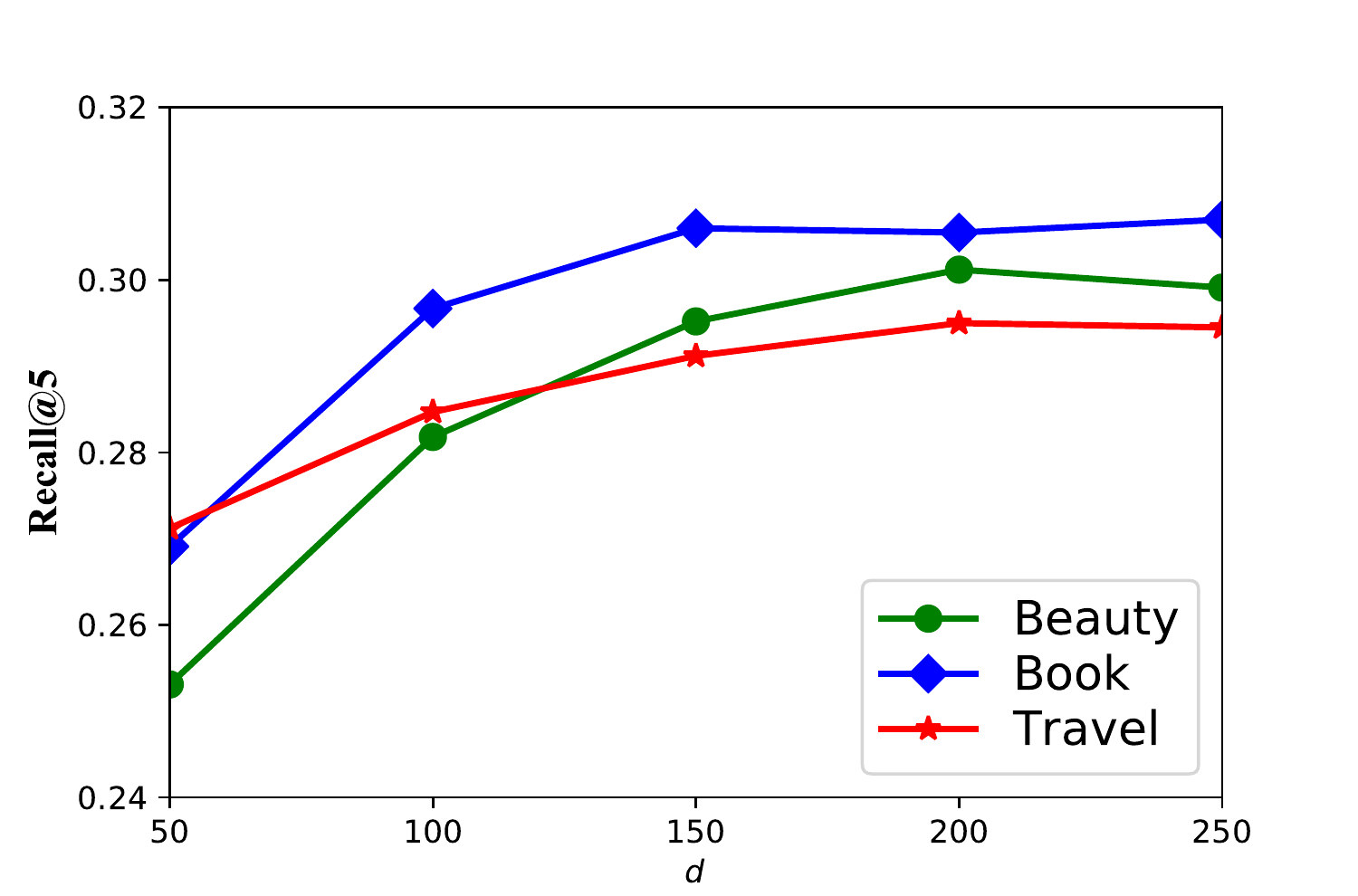} &
		\hspace{-0.2cm}\includegraphics[width=0.5\linewidth]{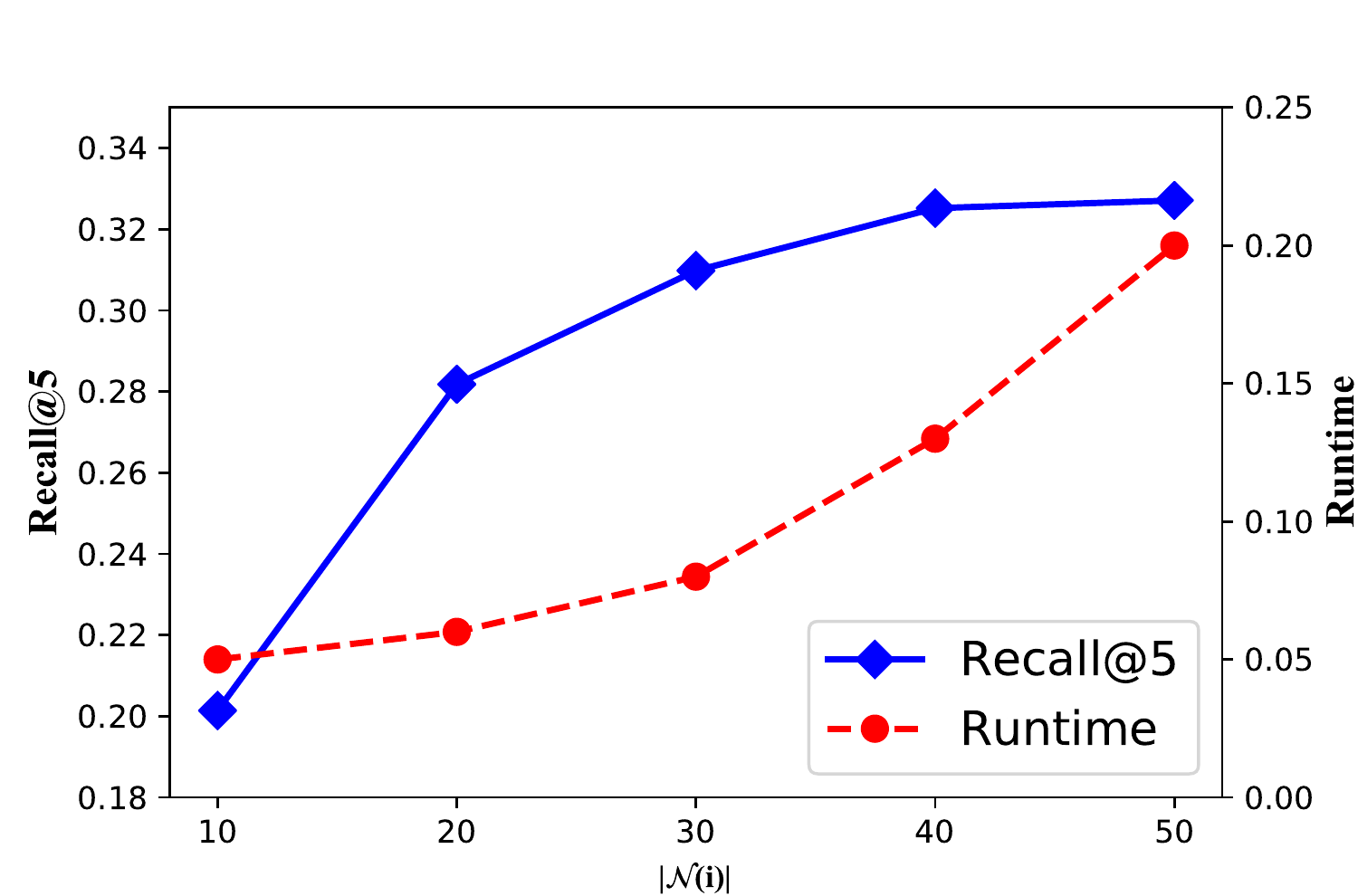}&\\
		
		\hspace{-0.2cm} \text{(a) dimension size $d$ } &
		\hspace{-0.2cm}\text{(b) neighbors sample size $|\mathcal{N}_{i}|$}&\\
		\end {array}$
		\vspace{-1mm}
		\caption{Parameter Sensitivity w.r.t the dimension of embedding $d$ and size of sample neighbors $|\mathcal{N}_{i}|$}\label{param_sensitity}
		\vspace{-4mm}
	\end{center}
\end{figure}

\textbf{Impact of the dimension size $d$}: We vary the dimension size of embedding from 50 to 250 by in increment of 50, and present the experimental results on multiple behavior tasks in Figure ~\ref{param_sensitity}(a). From the figure, we can observe that the performance raises when the size $d$ of dimension increases. This is because more dimensions can encode more useful information. However, the performance decreases when the dimension size $d_{k}$ continuously increase. The reason is that too large size of dimensions may introduce more spareness and noises which will reduce the performance.

\textbf{Impact of the sampling neighbors size $\mathcal{N}_{i}$}: In order to conduct this experiment, we fix the number of sampling neighbors $N_{i}$ in each layer to the same size, and change the number from 10 to 50. Then we illustrate the performance and runtime on beauty task in Figure~\ref{param_sensitity}(b). As the number of sampling neighbors increases, the margin of experimental performance gradually decreases, and the runtime of algorithm increases rapidly. Therefore, we often set the number of sampling neighbors as 20, which can balance the performance and runtime of MCNE well. 

\section{Conclusion}\label{conclusion}
In this paper, we proposed an end-to-end framework named MCNE to learn multiple conditional network representations for each user of social network. We designed a novel binary mask layer to separate the single node embedding to multiple conditional node embeddings for different behaviors. Then we introduced attention network to model the complex interactions among users' multiple aspects of similarity. And we proposed the adapted multi-aspect similarity message sending and receiving operation, in order to aggregate multi-aspect preference information from high-order neighbors. Furthermore, we utilized the Bayesian Personalized Ranking loss function to learn the users' preference similarity on each behavior, and jointly learn the multiple conditional node embeddings through multi-task learning framework. Compared with the-state-of-art baselines, MCNE can not only achieve significant performance improvements, but also illustrate the interpretability and robustness of learned conditional representations. In further work, we will attempt to combine the nodes' attribute information to improve the performance and interpretability of MCNE.

\begin{acks}
This research was partially supported by grants from the National Key Research and Development Program of China (No. 2016YFB1000904), the National Natural Science Foundation of China (Grants No. U1605251, 61703386), and the Anhui Provincial Natural Science Foundation (Grant No. 1708085QF140). Hao Wang and Qi Liu also thank the support of Tencent Rhino-Bird Joint Research Program.
\end{acks}

\bibliographystyle{ACM-Reference-Format}
\balance
\bibliography{sample-base}

%

%

\end{document}